\documentclass{emulateapj}\usepackage{apjfonts}

\usepackage{morefloats,url}
\usepackage{graphicx}
\usepackage{natbib}
\usepackage{amssymb}
\usepackage{xfrac}
\usepackage[hyperfootnotes=false]{hyperref}

\shorttitle{On the formation of eccentric millisecond pulsars}
\shortauthors{John Antoniadis}

\slugcomment{To appear in The Astrophysical Journal Letters}

\begin{document}

\title{On the formation of eccentric millisecond pulsars with helium white-dwarf companions}

\author{John Antoniadis\altaffilmark{1} }
\affil{Max-Planck-Institut f\"{u}r Radioastronomie, Auf dem H\"{u}gel 69, 53121 Bonn, Germany and \\
Dunlap Institute for Astronomy and Astrophysics, University of Toronto, 50 St. George Street
Toronto, Ontario M5S 3H4, 
Canada ; antoniadis@dunlap.utoronto.ca}
\received{2014 October 15}
\accepted{2014 November 15}
\altaffiltext{1}{Dunlap Fellow}

\begin{abstract}
Millisecond pulsars (MSPs) orbiting helium white dwarfs (WDs) in eccentric orbits 
challenge the established binary-evolution paradigm that predicts efficient  orbital  circularization  during the mass-transfer episode that spins up the pulsar. 
\cite{ft14} recently proposed  that these binary MSPs may instead form from the rotationally delayed accretion-induced collapse of a massive WD. 
This scenario predicts that eccentric systems preferably host low-mass pulsars and travel with small systemic velocities---in tension with new observational constraints. 
Here I show that a substantial growth in eccentricity may alternatively arise from the dynamical interaction of the binary with a circumbinary disk.  
Such a disk may form from ejected donor material during hydrogen flash episodes, when the neutron star is already an active radio pulsar and tidal forces can no longer circularize the binary. 
I demonstrate that a short-lived ($10^{4}-10^{5}$\,yr) disk can  result in eccentricities of $e\simeq 0.01-0.15$ for orbital periods between 15 and 50 days.
Finally, I propose that, more generally, the disk hypothesis  may explain  the lack of
circular binary pulsars for the aforementioned orbital-period range. 
\end{abstract}

\section{Introduction} 
The majority of binary millisecond pulsars (MSPs) orbit low-mass helium white-dwarf companions \citep[He\,WDs; see data in][]{mhth05}. The 
standard formation channel for these systems requires mass transfer from a low-mass donor ($m_{\rm donor} \lesssim 
2$\,M$_{\odot}$) onto an old neutron star (NS) during a long-lived, low-mass {X}-ray binary (LMXB) phase. Throughout the mass-transfer 
episode, the NS gains mass and spin angular momentum, and tidal forces acting on the donor wipe out any 
primordial eccentricity imparted on the system during the supernova explosion \citep[see, e.g.,][]{bvh91}.  

LMXBs with relatively wide initial separations  initiate mass transfer as the donor  undergoes hydrogen-shell burning on the red giant branch. Such systems evolve toward orbital periods between $\sim 2$ and $\sim 150$ days \citep{tv06} and follow two distinct relations.  
(1) the remnant companions are He\,WDs with masses that scale with orbital period, because the progenitor's core mass depends on the stellar radius which is set by the orbital separation \citep[e.g.,][]{sav87,ts99a}; (2) the residual eccentricity correlates with the orbital period  because turbulent density fluctuations in the donor's convective envelope prevent  perfect circularization \citep{phi92}. 

Indeed, most known binary MSPs outside of globular clusters closely follow the predictions of the  
``recycling'' scenario \citep[based on data available in the ATNF pulsar catalog;][]{mhth05}. Until recently, the only known MSP with a substantial eccentricity was PSR\,J1903+0327 with  $e\simeq0.44$ 
and a main-sequence companion \citep{crl+08}. The current 
consensus for this binary's formation is that it is the fossil of a triple system, where the mass-losing star was ejected due to dynamical interactions \citep[e.g.,][]{fbw+11,pzvh+11}. 
 
 Interestingly, eccentric MSPs have now grown in number with the discovery of the three systems shown in Table\,1. 
 For these binaries a triple-origin scenario seems unlikely for two main
 reasons: First, the companions are  He WDs (J.~Antoniadis {\it et al.}, in preparation) that have masses consistent with what is expected from binary evolution \citep[e.g.][]{ts99a}. Second, their orbital periods and eccentricities seem to resemble 
 each other, which is  unlikely  for systems that evolved through a chaotic dynamical process. 

\cite{ft14} proposed instead that the first two MSPs in Table\,1 may have formed directly from the rotationally delayed, accretion-induced collapse (RD-AIC) of a massive WD. In this scenario, the eccentricity grows from zero to roughly ${(m_{\rm WD}^{\rm progenitor} - m_{\rm NS})}/{M}\simeq 0.1$ during the NS formation (where $m_{\rm WD}^{\rm progenitor}$ is the mass of the WD progenitor, $m_{\rm NS}$ is the NS mass after the implosion, and  $M$ is the total mass of the binary). Systems formed via an RD-AIC should host low-mass NSs 
(with a mass equal to the initial baryonic WD mass minus the gravitational binding energy) and  move with small systemic velocities, as the natal kick from the WD implosion should be minimal \citep{kjh06,dbo+06}. 

However, recent observations show that eccentric MSPs in fact host NSs  with highly scattered masses and  systemic velocities, thereby  contradicting the RD-AIC hypothesis (J.~Antoniadis {\it et al.}, in preparation; E. Barr \& P. Freire, {private communication}). An explanation of these measurements within the RD-AIC scenario  would require an additional assumption of  highly differentially rotating WD progenitors and substantial natal kicks at birth \citep{ft14}. 

\begin{table}
\caption{Properties of Known Eccentric MSPs with Low-mass Companions}
\begin{center}
\begin{tabular}{cccc}
\hline
Name & Period (days) & Eccentricity &$m_{\rm WD, median}^{\rm a}$\\
\hline
J2234$+$0611$^{\rm b}$ & 32.01 & 0.129 & 0.23\\
J1946+3417$^{\rm c}$ & 27.01 & 0.134 & 0.24 \\
J1618$-$3919$^{\rm d}$ & 22.80 & 0.027 & 0.20 \\
\hline
\end{tabular}
\end{center}
\begin{flushleft}
$^{\rm a}$Assuming a pulsar mass of $1.4$\,M$_{\odot}$
and an inclination of $60^{\rm o}$. \\
$^{\rm b}$\cite{dsm+13}; $^{\rm c}$\cite{bck+13,ewan}; $^{\rm d}$\cite{eb01b,bai10}
\end{flushleft}
\end{table}

 In this Letter, I propose that high eccentricities may instead result from the dynamical interaction of the binary with a circumbinary disk (CB disk). I show that such a disk may form from material escaping the donor's surface as it undergoes  hydrogen-shell flashes,  following the LMXB phase and shortly before entering the final WD cooling branch.
The mass loss (of order $10^{-4}-10^{-3}$\,M$_{\odot}$) during flashes---occurring when the NS is already an active MSP and tidal forces can no longer affect the contracting proto-WD---can lead to a CB disk which may increase the eccentricity to the observed values within a timescale much shorter than the characteristic ages of these systems. Finally, using a Monte-Carlo (MC) simulation, I demonstrate that the CB hypothesis may explain  the lack of circular  MSPs with orbital periods between $\sim 15$ and $\sim 50$ days.  Within the CB disk scenario, the low end of the orbital period gap would be linked to a decreased CB disk lifetime for shorter orbital periods and the long-period cut-off would result from the cessation of hydrogen flashes for higher-mass WDs.
  
\section{Eccentricity growth though interaction with a circumbinary disk}
\subsection{Circumbinary disk formation, mass and lifetime}
A CB disk able to pump up the eccentricity of a binary MSP can form if the  following conditions are met.
\begin{itemize}
\item There should be enough matter lost from the donor that cannot be accreted onto the NS and
\item tidal forces that tend to circularize the orbit should be minimal.
\end{itemize}

Both these criteria are fulfilled immediately after the cessation of the long-term recycling phase. 
To demonstrate this I use a set of  LMXB  calculations conducted with the \textsc{mesa} stellar-evolution code \citep{mesa,mesa2}. 
The initial binaries consist of a 1.3\,M$_{\odot}$ NS (treated as a point mass), companions with masses between 1.4 and 1.6\,M$_{\odot}$ and initial orbital periods between two and four days. 
An example track for a donor mass of $m_{\rm donor} = 1.4$\,M$_{\odot}$ and an initial orbital period of $P_{\rm b}^{\rm init} = 3.0$\,days can be seen in Fig.\,1.
As a result of the hydrogen shell flashes, the binary undergoes 
a short-lived ($\sim10^3$\,yr) episode of additional Roche-lobe
overflow in which the mass-transfer rate exceeds the Eddington
limit of the accreting NS, causing matter to be lost from
the binary. The total ejected mass for the model system in Fig.~1 is $\Delta M_{\rm flash} \simeq 2.0\times 10^{-4}$\,M$_{\odot}$ and varies from $\sim 1\times10^{-4}$ to $9\times 10^{-4}$\,M$_{\odot}$ among different tracks \citep[see also][]{itl14, itl+14}. 

Immediately after the flash episode, the circularization timescale, $\tau_{\rm circ}$ 
\citep{zah77},  grows beyond $10^{10}$\,yr  as the star contracts and settles on the cooling track. Hence, tides become irrelevant for the subsequent evolution of the orbital elements. 
Furthermore, as previous studies have shown \citep[e.g.,][]{bdb02,dvb+06,tau12}, the NS is already an active MSP and its magneto-dipole pressure exceeds the ram pressure of the accreted matter at L1. Consequently, a 
large fraction of companion's material carrying substantial specific orbital angular momentum may escape the system through L2 and form a CB disk. 

\begin{figure}[h]
\begin{center}
\includegraphics[scale=0.4]{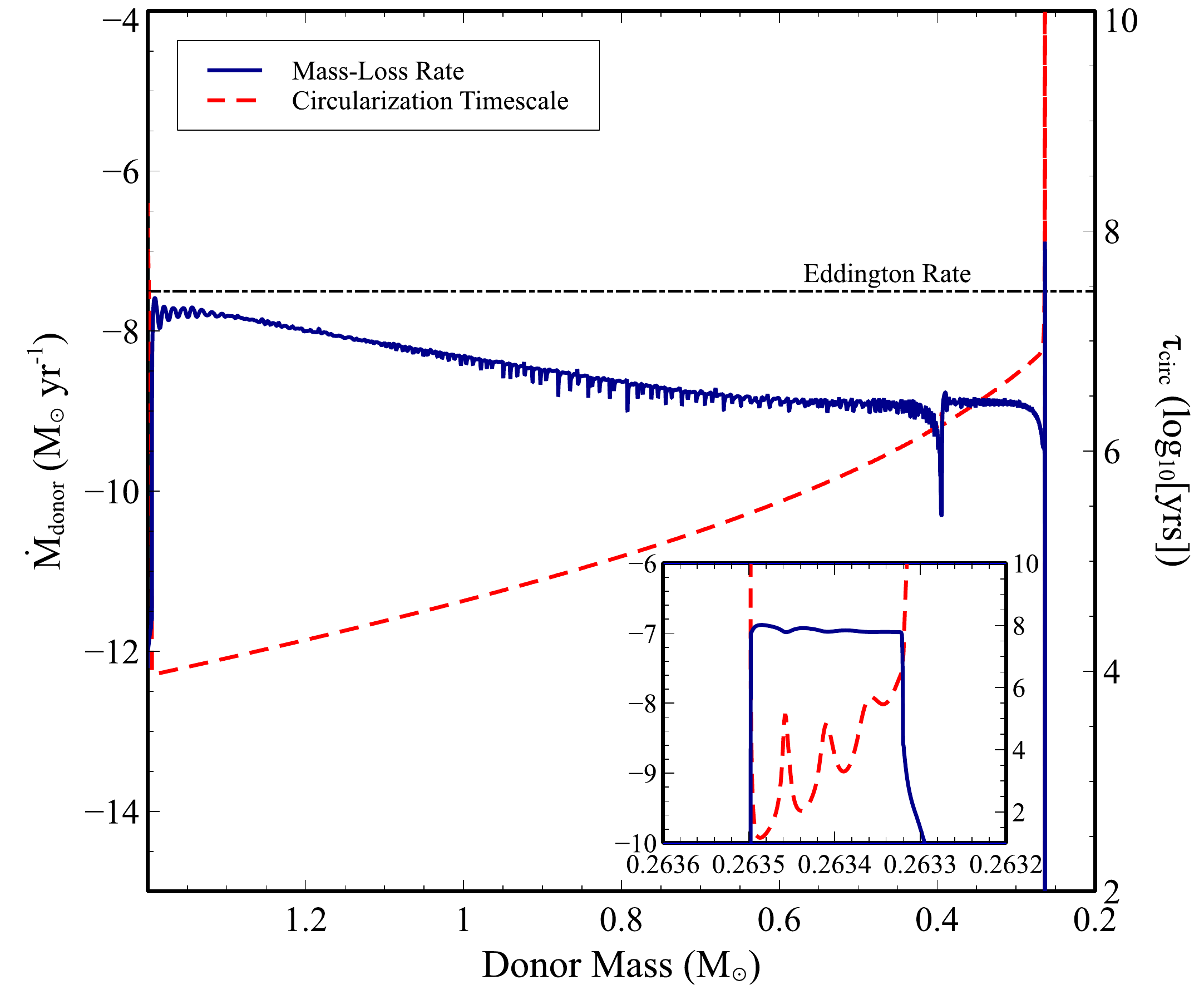}
\caption{Mass-transfer rate as a function of donor mass for an LMXB with  $m_{\rm donor}=1.4$\,M$_{\odot}$ and $P_{\rm b}^{\rm init} =3.0$\,days (see $\S 2.1$). The final period of the binary 
is $P_{\rm b}^{\rm final}\simeq 27$\,days. The red line shows the circularization timescale \citep{zah77} and the dashed line corresponds to the neutron-star Eddington accretion rate. The embedded figure displays in more detail the mass-transfer profile during the hydrogen-flash episodes of the donor. }
\label{figure:1}
\end{center}
\end{figure}

The lifetime of the disk is difficult to estimate but it most likely depends sensitively on the photo-evaporation efficiency of the pulsar's spin-down luminosity and the average distance from the 
binary \citep[and consequently the binary separation;][]{acp06,oce12,che13}. 
As we shall see below, for a  mean disk mass of $1.5\times 10^{-4}$\,M$_{\odot}$ throughout the evolution --- adopted here as a conservative estimate based on the binary evolution tracks --- the timescale to pump up
 the eccentricity  to $e=0.1$ is of the order of $10^4 - 10^5$\,yr, depending on the orbital period. Hence, for all following calculations, I take $\tau_{\rm CB,150}^{\rm max}=10^5$\,yr as an upper bound for a disk around 
a binary with a separation of $150$\,ls. This corresponds to an  average mass-loss rate of $\dot{M}_{\rm CB}\simeq10^{-9}$\,M$_{\odot}$\,yr$^{-1}$. Furthermore, as 
photo-evaporation  should also depend on the distance to the pulsar, I assume that the disk lifetime scales with the binary's semi-major axis as $\tau_{\rm CB}^{\rm max}(a) = \tau_{\rm CB,150}^{\rm max}(\frac{a}{150\,\rm{ls}})^2$.

\subsection{Impact of the Circumbinary Disk on the Orbital Parameters} 
The influence of a CB disk on the binary orbit has been studied extensively, especially in the context of eccentric post-asymptotic giant branch stars \citep{acl91,la96,la00a,dijv13}. 
Here, I adopt the  approach of \cite{dijv13} which is  based on the smooth-particle hydrodynamics simulations of \cite{la00a}. 
In this model, the resonant interactions between the binary and the disk are described using a linear perturbation theory, where the binary potential 
is approximated with the series expansion: 

\begin{equation}
\Phi(r,\theta,t) = \sum_{m,l} \phi_{m,l}(r) \exp[i m(\theta-(l/m)\Omega_{\rm b}t)].
\end{equation} 
Here, $l,m$ are integers and $\Omega_{\rm b} = 2\pi/P_{\rm b} = \sqrt{ GM/a^3}$ is the angular orbital frequency.
The disk is assumed to be thin and the kinematic viscosity is taken to be density independent. 
The inner edge of the disk is equal to the distance at which the resonant torque balances the viscous torque of the disk. 

For small eccentricities ($\lesssim0.2$), the rate of change for the orbital separation is given by \citep{dijv13}
\begin{equation}
\frac{\dot{a}}{a} = -\frac{2l}{m}\frac{M_{\rm disk}}{\mu}\alpha \left(\frac{H}{R}\right)^2 \frac{a}{R}\Omega_{\rm b},
\end{equation}
where $\mu$ is the reduced mass of the binary, $R$ is the half orbital angular-momentum radius of the disk, $H/R$ the disk thickness (here fixed to $H/R = 0.1$), and $\alpha (= 0.1)$  the viscosity parameter.
The change in eccentricity is given by
\begin{equation}
\dot{e} = \frac{2(1-e^2)}{e + \frac{\alpha}{100e}}\left(\frac{l}{m} - \frac{1}{\sqrt{1-e^2}}\right)\frac{\dot{a}}{a}.
\end{equation}
For $e \lesssim 0.1\alpha^2$, the rate of eccentricity change is primarily determined by the $l=1,m=1$ resonance, while for large values by the $l=1,m=2$ resonance 
\citep{dijv13,la00a}. 
\begin{figure}[h]
\begin{center}
\includegraphics[scale=0.5]{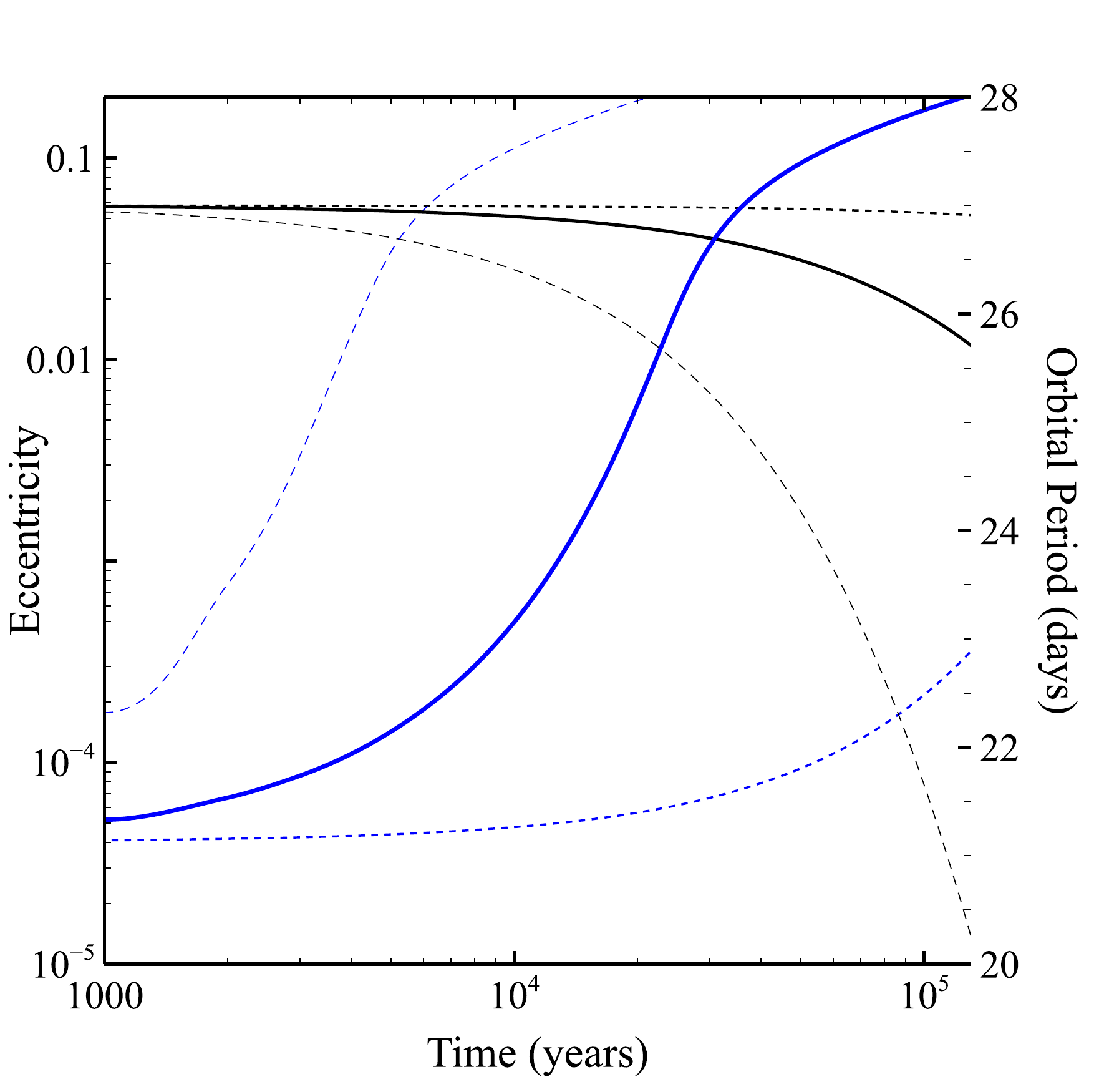}
\caption{Change of the orbital elements due to the interaction of a binary with a CB disk based on eqs.\,1\& 2. The binary shown here has $P_{\rm b}^{\rm init}= 27$\,days, $e_{0}=4\times10^{-4}$ \citep{phi92}, $m_{\rm PSR}= 1.45$\,M$_{\odot}$ and $m_{\rm WD}=0.281$\,M$_{\odot}$ \citep{ts99a}. The blue lines depict the evolution of eccentricity while the black lines the evolution of orbital period. Solid, dotted and dashed lines represent calculations for disk masses of $1.5\times10^{-4}$, $10^{-5}$ and $10^{-3}$\,M$_{\odot}$ respectively. }
\label{figure:2}
\end{center}
\end{figure}
Using  Eqs.~(2)~and~3, one finds that for a typical binary MSP with $\mu = 0.24$ and $P_{\rm b} = 27$\,days, and for a constant disk mass of $1.5\,\times10^{-4}$\,M$_{\odot}$ (see above), the time needed to excite the eccentricity to $e\simeq0.1$ is only $\sim 
50\,000$\,yr, therefore providing a possible explanation for the origin of eccentric MSPs (see Fig.~2).

\subsection{Monte-Carlo Simulations and the Origin of the Orbital Period Gap}
To assess whether the CB disk scenario can reproduce the observed eccentricities, I perform a Monte Carlo (MC) simulation under certain assumptions for the disk lifetime, distribution of orbital periods and stellar masses.

For the MC, Eqs.~(2) and~(3) are solved 10\,000 times assuming a pulsar mass drawn from a normal distribution with a mean of 1.45\,M$_{\odot}$ and $\sigma=0.2$\,M$_{\odot}$,  and a disk mass of $M_{\rm disk}=1.5\times 10^{-4}$\,M$_{\odot}$ (see $\S 2.1$). 
The orbital period distribution is taken to be flat in $\log (P_{\rm b})$ between 1 and 50\,days, which roughly corresponds to He\,WD masses within the critical range for hydrogen flashes (see $\S 3$ for discussion). $m_{\rm WD}$ is calculated using the \cite{ts99a} relation for solar metallicities ($Z=0.02$). 

For the maximum disk lifetime, I adopt  $\tau_{\rm CB, 150}^{\rm max} = 10^5$\,yr, which is required to explain the eccentricities of the systems shown in Table\,1 (see also $\S2.1$).
To simulate the effect of a varying photo-evaporating efficiency, for each orbital separation $a$, the disk  lifetime is conservatively assumed to follow  a  flat distribution between 0 and $\tau_{\rm CB, 150}^{\rm max}\left( \frac{a}{150\,{\rm ls}} \right)^2$ (see $\S2.1$). 

Finally, the initial binary eccentricity before the CB disk formation is fixed to the \cite{phi92} prediction with a spread following a Boltzmann distribution with $\sigma(e_{0}^2) = e_{0}^2/2$. 

The main results of the MC study are discussed in the following section.

\section{Results}
\begin{figure*}[t]
\begin{center}
\includegraphics[scale=0.28]{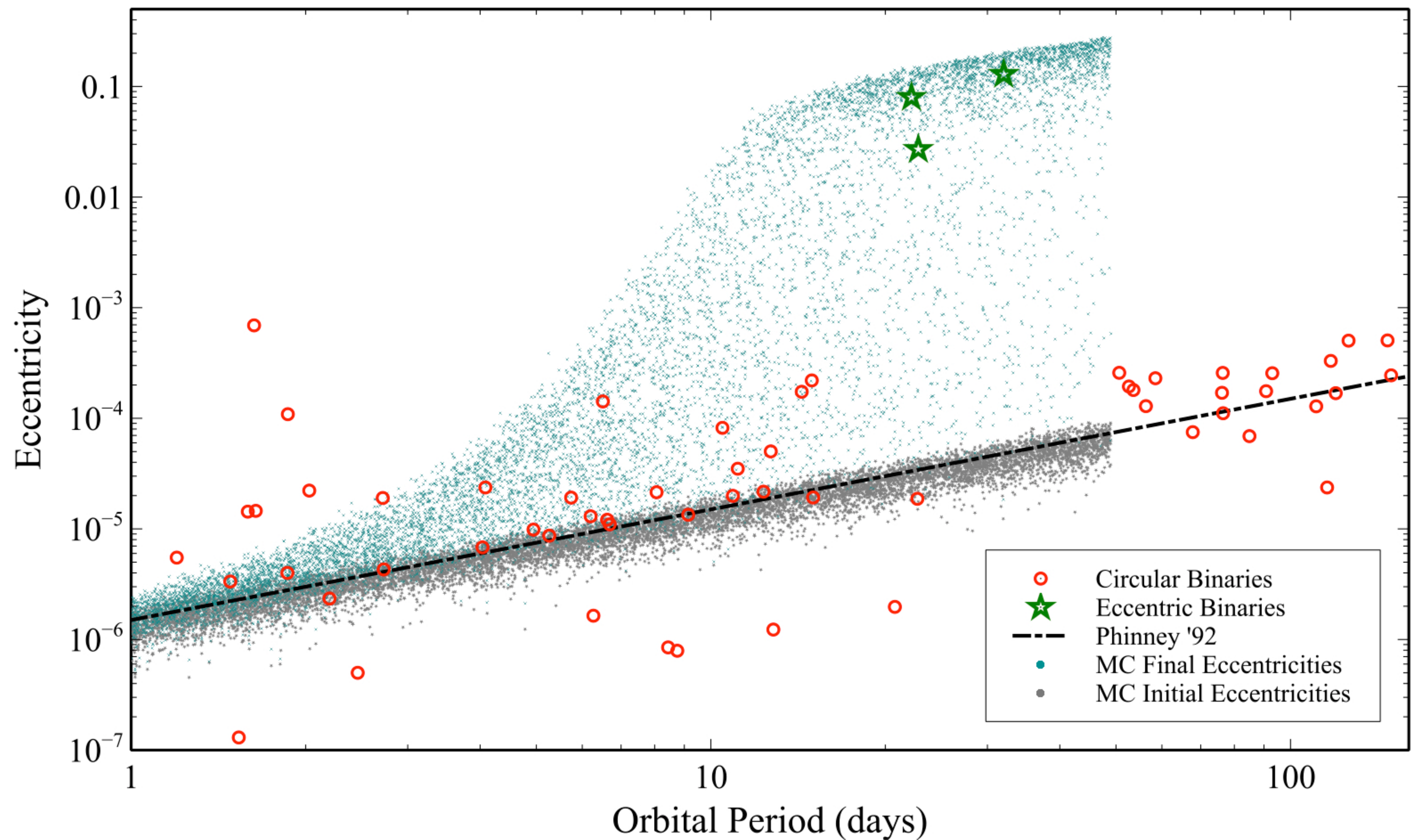}
\caption{Eccentricities of Galactic field binary MSPs  as a function of their orbital periods \citep[data taken from][]{mhth05,dsm+13,bck+13,eb01b,bai10}. The dark blue points are the results of the MC simulation based on the CB disk scenario (see the text). Black points show the distribution of eccentricities expected from \cite{phi92}. }
\label{figure:2}
\end{center}
\end{figure*}
Fig.\,3 shows a qualitative comparison between the observed eccentricities of MSPs with He\,WD companions and the results of the MC simulation described in $\S2$. 
The most interesting feature of the model distribution is a ``jump'' in the orbital eccentricities for systems with $P_{\rm b} \geq 10$\,days.  This jump is a result of 
 higher initial eccentricities for these periods but also requires an increased disk lifetime (10${^4}-10^5$\,yr). Interestingly, as the number of circular binaries decreases significantly for longer orbital periods, the CB scenario provides a possible explanation for the under-abundance of circular MSPs with orbital periods between 15 and 50\,days  \citep{cam95}. 
If this is the case, the upper $P_{\rm b}$ limit of the gap ($\sim 50$\,days) should mark the critical He\,WD mass above which hydrogen flashes cease \citep[$\sim 0.31$\,M$_{\odot}$ based on][]{ts99a}. 

If the disk lifetime is independent of the distance to the pulsar,  the orbital period gap should correspond exactly to the mass range for 
which flashes occur. However, this would not explain the observed abundance of systems at $P_{\rm b}\leq 14$\,days with eccentricities larger than the \cite{phi92} prediction. Furthermore, the corresponding lower critical WD mass for flashes ($\sim 0.26$\,M$_{\odot}$) is far too high compared to current model predictions \citep[e.g. see:][and references therein]{amc13}. 
Finally, note that, albeit less frequent, binaries with $e \leq 10^{-3}$ can still exist inside the gap, although preferably for small disk lifetimes compared to the maximum value. This may be the case for  e.g. MSPs with high spin-down luminosities and/or favorable geometrical alignment \citep{ccth13,gt14}. 

Despite the qualitative agreement between the observed and simulated populations, it should be noted that some of the critical features described above  
depend sensitively on physical parameters that are not yet well constrained observationally. For example, the maximum eccentricity for a given orbital period would 
change significantly if one varies the disk mass and lifetime (here chosen to fit the existing data, see also Fig\,2). Similarly, a different distribution of disk lifetimes  
and/or  a change in the functional form describing the dependence between maximum lifetime and orbital period  would impact the relative fraction of eccentric MSPs 
and the critical orbital period for the transition from circular to eccentric systems. 
A final point of caution is the model describing the interaction with the CB disk. In particular, a feature of the prescription adopted here is the small growth in 
eccentricity at the limit $e\rightarrow0$, arising because the central cavity of the disk is circular.  However, recent high-resolution shock capturing simulations 
\citep[e.g.,][]{skl+12, nmn+12, dhm13} suggest that the central cavity may be highly eccentric, even for equal-mass binaries. This would cause eccentricity pumping 
even for the most circular binaries and remove the dependence to the distribution of initial eccentricities.

\section{Summary}
In this Letter, I demonstrate that  eccentric MSPs with He\,WD companions can form through the interaction of the binary with a CB disk fed by matter escaping the WD 
during hydrogen-flash episodes. Under this hypothesis, the recently discovered MSPs with $e\simeq 0.01-0.13$ require a CB disk interacting with the binary over 
$10^{4}-10^{5}$\,yr, for the typical mass loss ($\sim 1-9\times 10^{-4}$\,M$_{\odot}$) expected during the flash episodes. 
Based on the MC simulation, conducted here as a proof of concept, the CB disk scenario makes the following predictions: 
\begin{enumerate}
\item The companions of eccentric MSPs should be He\,WDs that follow the $P_{\rm b}$--$m_{\rm WD}$ relation for LMXB evolution.
\item Both MSP masses and systemic velocities should closely resemble those of circular binary MSPs.
\item The maximum $P_{\rm b}$ at which eccentric MSPs can be found should correspond to the maximum critical He\,WD mass for the occurrence of flashes.  
\item There should be a maximum value for the eccentricity, determined primarily by the (product of) CB disk mass and lifetime. 
\item Small eccentricities may still exist within the period gap, but preferably for binaries hosting highly energetic MSPs and/or high photo-evaporation efficiencies. 
\end{enumerate}

As some of these predictions depend sensitively on underlying assumptions that are loosely constrained by observations, a better sampling of the binary MSP population will provide valuable insights on the scenario and its parameters. 

\acknowledgments{I am grateful to Thomas Tauris, Paulo Freire and Robert Izzard for helpful discussions related to this work. I also thank the anonymous referees for their useful suggestions. The Dunlap Institute is funded by an endowment established by the David Dunlap family and the University of Toronto. This research was  funded by the European Research Council Grant BEACON, under contract no.\,279702 (PI: Paulo Freire). I have made extensive use of NASA's Astrophysics Data System.\\ All data and source code can be made available upon request.}

\end{document}